\documentstyle[aps,epsfig,multicol]{revtex}
\title{Electron Spins in Artificial Atoms and Molecules for
Quantum Computing~\footnote{Invited review prepared for Special Issue 
of Semiconductor Science and Technology, "Semiconductor Spintronics", 
ed. H. Ohno, 2002.}}

\author{Vitaly N. Golovach and Daniel Loss}

\address{Department of Physics and Astronomy, University of Basel,\\
 Klingelbergstrasse 82, CH-4056 Basel, Switzerland}

\date{\today}

\newcommand{\XOR}{\textsc{xor}}

\newcommand{\sqrtSwap}{{U_{\rm sw}^{1/2}}}

\newcommand{\bra}[1]{{\langle #1 |}}
\newcommand{\ket}[1]{{| #1 \rangle}}
\newcommand{\spup}{\ket{\!\uparrow}}
\newcommand{\spdown}{\ket{\!\downarrow}}

\newcommand{\qdot}[1]{\begin{picture}(13,10)
    \put(6,3.6){\circle{13}}
    \put(6,3.6){\makebox(0,0){#1}}
    \end{picture}}
\newcommand{\edot}[1]{\begin{picture}(7,9) 
    \put(3,3.6){\makebox(0,0){#1}}
    \end{picture}}

\begin{document}
\maketitle
\begin{abstract}
Achieving control over the electron spin in quantum dots (artificial atoms)
or real atoms promises access to new technologies in conventional
and in quantum information processing. 
Here we review our proposal for quantum
computing with spins of electrons confined to quantum dots.
We discuss the basic requirements for implementing spin-qubits, 
and describe a complete set of quantum gates for single- and two-qubit
operations.
We show how a quantum dot attached to leads can be used for spin filtering 
and spin read-out, and as a spin-memory device. 
Finally, we focus on
the experimental characterization of the quantum dot systems, and discuss
transport properties of a double-dot and show how Kondo correlations can be 
used to measure the Heisenberg exchange interaction between the spins of 
two dots.
\end{abstract}

\ifpreprintsty\else\begin{multicols}{2}\fi

                  \section{Introduction}                         %
\label{secIntro}
Coulomb blockade phenomena have atracted much interest during the last 
few dacades~\cite{kouwenhoven}. Creation of confined electron systems at the 
nanometer scale has made it possible to study the quantum-meachnical 
nature of the band electron in a variety of materials. The tunability 
of the quantum-dot devices provided a unique opportunity to study the 
charging effects, and hence, the correlation effects associated with 
the Coulomb charging energy. However, in confined 
systems of smaller sizes, where the size-quantization energy is resolved,
new correlations set in due to the Pauli exclusion principle. 
This brings along the electron spin as a degree of freedom in quantum
confined structures, and accessing it in a deterministic way would allow 
for novel implementations in quantum information processing.
An increasing number of spin-related
experiments~\cite{Prinz,Kikkawa,Fiederling,Ohno,Roukes,Ensslin} indeed
show that the spin of the electron in quantum-confined nanostructures 
is a promising candidate for information processing, 
due to the unusually long (100's nanosecs) spin dephasing 
times~\cite{Kikkawa,Fiederling,Ohno}.
On the other hand there are propsed methods~\cite{Loss97} for efficient 
and deterministic control of the spin state in single quantum dots 
as well as methods of entangling the spins of two dots, the latter being a 
crucial element in quantum information processing. 
Thus, the field of interest falls into two parts, 
one being improving the technologies for conventional computation, 
and the other -- implementing fundamentally new algorithms of computation 
with quantum bits of information (qubits) and devising a scalable quantum 
computer in the long run. 
In conventional computers, the
electron spin can be expected to enhance the performance of quantum
electronic devices, such as spin-transistors (based on
spin-currents and spin injection), non-volatile memories,
single spin as the ultimate limit of information storage 
etc.~\cite{Prinz,Recher}.
For implementing quantum computing~\cite{NC}, as first pointed
out in Ref.~\onlinecite{Loss97}, the spin of a confined electron 
appears as the most natural candidate for the qubit. 
Indeed, provided the spin-orbit coupling is negligible, 
the intrinsic two-state space of the spin encodes exactly one qubit 
and allows for no undesired parts of the Hilbert space, transitions to 
which could lead to leakage errors in quantum computation.
We have shown~\cite{Loss97} that
the spin qubits, when located in quantum-confined structures such
as semiconductor quantum dots or atoms or molecules, satisfy all
requirements needed for a scalable quantum computer.

The long distances, of up to $100\:\mu{\rm m}$~\cite{Kikkawa}, over which 
spins can be transported phase-coherently, make the electron spin a 
plausible candidate for quantum information transmission in solid state
devices. A spin-qubit attached to a mobile electron can be transported 
along conducting wires between different subunits in a quantum 
network~\cite{MMM2000,BLS}. 
Entangled electrons, which can be created in coupled
quantum dots or via a superconductor~\cite{Recher1}, 
provide a source of Einstein-Podolsky-Rosen (EPR) 
pairs~\cite{MMM2000,BLS}, wich are necessary for secure quantum communication.

The methods used to implement the electron spin in conventional computers 
and in quantum computers are often identical, because of the 
quantum-mechanical nature of the electron and its spin. Our short-term goal 
is to find ways to control the coherent dynamics of electron spins in 
quantum-confined nanostructures. The use of solid state physics as a base for
implementing the quantum computer is motivated by the unparalleled 
flexibility in designing an appropriate medium for the realization of
a given physical phenomena.  

In the following, we review the status of our theoretical
efforts towards the goal of implementing quantum
computation with electron spins in
quantum-confined nanostructures.

\subsection{Quantum Computing and Quantum Dots}
\label{ssecQC}
The possibility of outperforming classical computation, 
which opens up in quantum algorithms such 
as the one discovered by Shor~\cite{Shor94} and by Grover~\cite{Grover},
has attracted much interest.
A quantum algorithm makes use of the quantum computers's ability
to exist in any superposition of the states of 
its binary basis and to perform quantum time evolution 
for computation; hence the parallelism of quantum computing.
The requirement for the quantum bit of information (qubit), 
which is at the heart of the quantum computer, 
is that it can exist in any state of a quantum two two-level system,
{\em i.e.} $|\psi\rangle=\alpha|0\rangle+\beta|1\rangle$, where
$|0\rangle$ and $|1\rangle$ are the states of the ``classical'' bit,
and $|\alpha|^2+|\beta|^2=1$.
Apart from this, a qubit should be able to couple to any other
qubit in the quantum comupter and form a coherent two-qubit state.
These two elements, which are the one- and two-qubit gates, are sufficient 
for forming a many-qubit coherent state and implementing any quantum algorithm.

A recently growing list of quantum tasks~\cite{MMM2000,Bennett00} such as
cryptography, error
 correcting schemes,  quantum teleportation, etc. 
have indicated even more the desirability of experimental
implementations of quantum computing.
On the other hand, there is also
a growing number of proposed physical
 implementations of qubits and quantum gates. A few examples are:
 Trapped ions~\cite{traps},
 cavity QED~\cite{cavity},
 nuclear spins~\cite{nmr,Kane},
 superconducting devices~\cite{Schon,Averin,Ioffe,Mooij},
 and our qubit proposal~\cite{Loss97} based on the spin of the electron
 in quantum-confined nanostructures, and in particular in quantum dots
with an all-electrical control of spin. Subsequent proposals such 
as~\cite{Kane,Privman,Yablonovitch} are based on the same principles as 
introduced in~\cite{Loss97} and reviewed herein.

\ifpreprintsty\else\end{multicols}\fi
\begin{figure}[t]
\centerline{\psfig{file=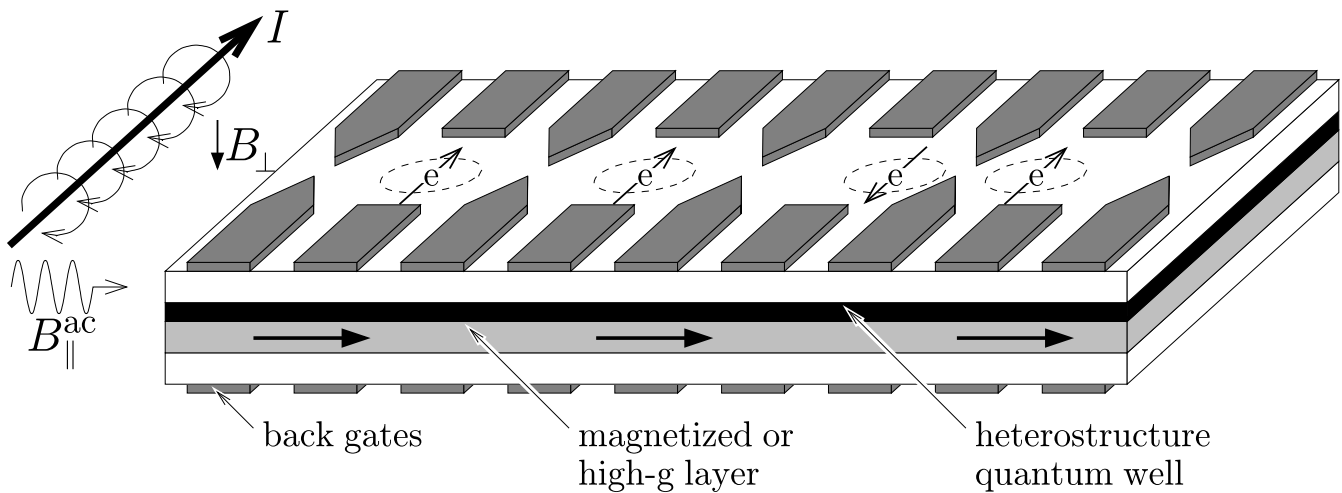,width=17.8cm}}
\caption{\label{figArray}
Quantum dot array, controlled by electrical gating.
The electrodes (dark gray) define quantum dots (circles) by confining 
electrons.
The spin 1/2 ground state (arrow) of the dot represents the qubit.
These electrons can be moved by electrical gating into the magnetized or 
high-$g$ layer, producing locally different Zeeman splittings.
Alternatively, magnetic field gradients can be applied, as e.g. produced
by a current wire (indicated on the left of the dot-array).
Then, since every dot-spin is subjected to a different Zeeman
splitting, the spins can be addressed individually, e.g. 
through ESR pulses of an additional in-plane magnetic ac field
with the corresponding
Larmor frequency $\omega_{\rm L}=g\mu_{B}B_\perp/\hbar$.
Such mechanisms can be used for single-spin rotations and the
initialization step.
The exchange coupling between the quantum dots can be controlled by
lowering the tunnel barrier between the dots.
In this figure, the two rightmost dots are drawn schematically as 
tunnel-coupled. Such an exchange mechanism can be used for the XOR
gate operation involving two nearest neighbor qubits. The XOR operation
between distant qubits is achieved by swapping (via exchange)  the qubits
first to a  nearest neighbor position. 
The read-out of the spin state can be achieved via 
spin-dependent tunneling and SET devices~\protect\cite{Loss97}, 
or via a transport current passing the dot~\protect\cite{EL}.
Note that all spin operations, single and two
spin operations, and spin read-out, are controlled electrically via the
charge of the electron and not via the magnetic moment of the spin. Thus, no
control of local magnetic fields is required, and the spin is only used for
storing the information.
This spin-to-charge conversion is based on the Pauli principle and Coulomb
interaction and allows for very fast switching times (typically picoseconds). 
A further advantage of this all-electrical scheme is its scalability into an array of 
arbitrary size. }
\end{figure}
\clearpage
\ifpreprintsty\else\begin{multicols}{2}\fi

Semiconductor quantum dots are structures where charge carriers
 are confined in all three spatial dimensions,
 the dot size being of the order of the Fermi wavelength
 in the host material,
 typically between $10\:{\rm nm}$ and $1\:{\rm \mu m}$~\cite{kouwenhoven}.
The confinement is usually achieved by electrical gating of a
 two-dimensional electron gas (2DEG),
 possibly combined with etching techniques, see Fig.~\ref{figArray}.
Precise control of the number of electrons in the 
conduction band
 of a quantum dot (starting from zero) has been achieved in GaAs
 heterostructures~\cite{tarucha}.
The electronic spectrum of typical quantum dots can vary strongly
 when an external magnetic field is applied~\cite{kouwenhoven,tarucha},
 since the magnetic length corresponding to typical laboratory fields
 $B\approx 1\,{\rm T}$ is comparable to typical dot sizes.
In coupled quantum dots
 Coulomb blockade effects~\cite{waugh},
 tunneling between neighboring dots~\cite{kouwenhoven,waugh},
 and magnetization~\cite{oosterkamp} have been observed as well as the
 formation of a delocalized single-particle state~\cite{blick}.

\section{General Considerations for Quantum Computing with Spins}%
\label{secGeneral}

\subsection{Coherence}
\label{ssecCoherence}
Magneto-optical experiments,
based on time-resolved Faraday rotation
measurements, show long spin coherence times
in doped GaAs in the bulk and a 2DEG\cite{Kikkawa}.
At $B=0$ and $T=5\:{\rm K}$, a transverse spin lifetime
(dephasing time) $T_2^*$ exceeding $100\:{\rm ns}$ was measured,
with experimental indications that this time is a single-spin
effect\cite{Kikkawa}.
Since this number still includes inhomogeneous effects---e.g.\ 
g-factor variations in the material,
leading to spins rotating with slightly different frequencies
and thus reducing the total magnetization---it represents 
only a lower bound of the decoherence time $T_2$
of a {\it single} spin, $T_2\ge T_2^*$,
which is relevant for using spins as qubits.
Using the same pump-probe technique, spin dephasing
times in semiconductor (CdSe) quantum dots have been measured~\cite{Gupta},
with at most one spin per dot.
The relatively small $T_2^*$ dephasing times (a few ns at vanishing
magnetic field), which have been seen in these experiments,
probably originate from a large inhomogeneous
broadening due to a strong variation of g-factors~\cite{Gupta}.
Nevertheless, the fact that many coherent
oscillations were observed~\onlinecite{Gupta}
provides strong experimental support to the idea of using
electron spin as a qubit.
 
\subsection{Upscaling}
\label{ssecScaling}
To outperform a classical computer, a quantum computer
will need a number of qubits on the order of $10^5$.
Hence, it is essential that the underlying concept
can be scaled up to a large number of qubits.
This scaling requirement is, in principle, achievable with
spin-based qubits confined in quantum dots,
since producing arrays of quantum dots~\cite{MMM2000,ankara}
is feasible with today's technology
of defining nanostructures in semiconductors.
Of course, the actual implementation of such arrays (see Fig.~\ref{figArray})
including all the needed circuits poses tremendous experimental challenges,
but at least we are not aware of any physical restriction which would
exclude such an upscaling for spin-qubits.

\subsection{Switching}
\label{ssecPulse}
Quantum gate operations can be controlled through an 
effective Hamiltonian (see Sec.~\ref{secSSpinRot} and~\ref{sec2bit})
\begin{equation}
\label{eqnGenH}
H(t) = \sum_{i<j} J_{ij}(t) \, {\bf S}_i\cdot{\bf S}_j
 + \sum_i \mu_B g_i(t)\, {\bf B}_i(t) \cdot {\bf S}_i
\,.
\end{equation}
The coupling constants $J_{ij}$, $g_i$ and the magnetic field (local)
${\bf B}_i$ are controlled via external gate fields, 
which are switched with some pulses $v(t)$.
In the following we assume $J_{ij}$ to be non-zero only for the
neighboring qubits. Note, however, that in cavity-QED systems there
is also a long-range coupling of qubits~\cite{Imamoglu}, and
that long-range coupling via a superconductor is also possible~\cite{CBL}.
But even if the exchange coupling is only local,
operations on non-neighboring qubits can still be performed. 
This is achieved by swapping states of neighboring qubits
(see Sec.~\ref{sec2bit}), which allows one to move the qubit
around in an array of quantum dots and couple it to the desired other qubit.

For the gating mechanisms described in Sec.~\ref{secSSpinRot} 
and~\ref{sec2bit}, only the time integral $\int_0^{\tau} P(v(t)) dt$ 
(mod $2\pi$) is important. 
Here, $P(v(t))$ stands for the exchange coupling $J$ or the Zeeman 
interaction. 
The requirement on the pulse shape is that it does not violate the
validity of the effective Hamiltonian (\ref{eqnGenH}), but otherwise the 
gating mechanisms are independent of the actual shape of $v(t)$. 
Since the effective Hamiltonian (\ref{eqnGenH}) was obtained by projecting 
out higher energy states of one and two coupled dots 
(see Sec.~\ref{ssecldots}), care should be taken
that the pulses do not excite the quantum dots to the projected-out
energy levels. This can be achieved by switching $v(t)$ adiabatically,  
i.e.\ such that $|\dot{v}/v| \ll \delta\varepsilon/\hbar$,
where $\delta\varepsilon$ is the energy scale on which excitations may
occur.
We find that $v(t) = v_0\,{\rm sech}(t/\Delta t)$, where $v_0$ is the pulse
amplitude and $\Delta t$ the characteristic width, is optimal for 
a fast adiabatic switching, provided 
$1/\Delta t \ll \delta\varepsilon/\hbar$.
For a detailed analysis of adiabatic switching, see~\cite{Schliemann}.

A single qubit operations can be performed
for example in g-factor-modulated materials,
as described in Sec.~\ref{secSSpinRot}.
A spin can be rotated by a relative angle of
$\varphi=\Delta g_{\rm eff} \mu_B B \tau/2\hbar$
through changing the effective g-factor by $\Delta g_{\rm eff}$
for a time $\tau$.
Thus, a typical switching time for an angle $\varphi=\pi/2$,
a field $B=1\:{\rm T}$,
and $\Delta g_{\rm eff}\approx 1$
is $\tau_s \approx 30\:{\rm ps}$.
If slower operations are required, they are easily implemented
by choosing a smaller $\Delta g_{\rm eff}$,
reducing the magnitude of the field $B$,
or by replacing $\varphi$ by $\varphi+2\pi n$ with integer $n$.

Next we consider two exchange-coupled spins,
 which perform a square-root-of-swap gate
 for the integrated pulse $\int_0^{\tau_s}J(t)dt/\hbar=\pi/2$,
 as described in Sec.~\ref{sec2bit}.
We apply a pulse
$J(t) = J_0\,{\rm sech}((t-\tau_s/2)/\Delta t)$ with
$J_0 = 80\:\mu{\rm eV}$, and
 choose $\Delta t = 4\:{\rm ps}$, which gives for the
switching time $\tau_s \approx 30{\rm ps}$, and
the adiabaticity criterion
 $\hbar/\Delta t \approx 150 \:\mu{\rm eV} \ll \delta\varepsilon$.

\subsection{Error Correction}
\label{ssecError}
Realization of a reliable error-correction scheme~\cite{errCorr} is
one of the main goals in quantum computation.
The known schemes for fault-tolerant quantum computation work if 
the gate operation error rate does not exceed a certain threshold value, 
usually about $10^{-4}$ (depending on the scheme)\cite{preskill}.
If we take the ratio of the switching times from Sec.~\ref{ssecPulse},
$\tau_s\approx 30\:{\rm ps}$, and
the dephasing time from Sec.~\ref{ssecCoherence},
$T_2\ge 100\:{\rm ns}$,
we obtain a value close to this threshold.
Thus, in our proposal, we can expect an arbitrary upscaling of the quantum 
computer, and we are
no further limited by decoherence and lacking gate precision.
We note that implementing an error-correction scheme requires
a larger number of gate operations, and therefore,
it is desirable to perform them in parallel;
otherwise the pursued gain in computational power is used up
for error correction.
Hence, one favors concepts
 where a localized control of the gates can be realized
 such that operations can be performed in parallel.
However, since there are still
 many milestones to reach
 before sophisticated error-correction schemes can be applied,
 one should by no means disregard setups
 where gate operations  are performed  in a serial way.

             \section{Single-Spin Rotations}                     %
\label{secSSpinRot}

For quantum computing it is necessary (but not sufficient) 
to perform one-qubit operations. In the context of spin-qubits,
it translates into single-spin rotations.
This can be achieved by exposing a specific qubit to a
time-varying Zeeman coupling
$(g\mu_B {\bf S}\cdot {\bf B})(t)$~\cite{BLD},
which can be controlled through
both the magnetic field ${\bf B}$ and/or the g-factor $g$.
Since only phases have a relevance,
it is sufficient to rotate all spins of the system at once
(e.g.\ by an external field $B$),
but with a different Larmor frequency.

Localized magnetic fields can be generated
with the magnetic tip of a scanning force microscope,
a magnetic disk writing head,
by placing the dots above a grid of current-carrying wires,
 or by placing a small wire coil above the dot etc.

Single-spin rotations can be achieved by ESR techniques~\cite{BLD}.
One applies a static local magnetic field $B$ for the qubit(s), 
which should be rotated. An ac magnetic field is then applied 
perpendicular to the first field with the resonant frequency that matches 
the Larmor frequency $\omega_L=g\mu_BB/\hbar$. 
Due to paramagnetic resonance~\cite{Shankar}, 
this causes spin-flips in the quantum dots with the corresponding 
Zeeman splitting. 

The equilibrium position of the electron
 can be moved around through electrical gating.
Thus, if the electron wave function is pushed into a region
 with a different magnetic field strength or (effective) g-factor,
one produces a relative rotation around the direction of ${\bf B}$ by an
 angle of $\varphi=(g'B'-gB)\mu_B\tau/2\hbar$, see Fig.~\ref{figArray}.
Regions with an increased magnetic field can be provided
by a magnetic (dot) material
while an effective magnetic field can be produced e.g.\ with
dynamically polarized nuclear spins (Overhauser effect)~\cite{BLD}.

We shall now explain a concept for using g-factor-modulated
 materials~\cite{MMM2000,ankara}.
In bulk semiconductors
 the free-electron value of the Land\'e g-factor $g_0=2.0023$
 is modified by spin-orbit coupling. Similarly, the g-factor
can be drastically enhanced by doping the semiconductor
with magnetic impurities~\cite{Ohno,Fiederling}.
In confined structures such as quantum wells, wires, and dots,
 the g-factor is further modified and becomes sensitive
 to an external bias voltage~\cite{Ivchenko}.
We have numerically analyzed a system with a layered structure
 (AlGaAs-GaAs-InAlGaAs-AlGaAs),
 in which the effective g-factor of electrons is varied by
 shifting their equilibrium position from one layer to
 another by electrical gating.
We have found that in this structure the effective g-factor
 can be changed by about $\Delta g_{\rm eff}\approx 1$~\cite{ankara}.
Such a gate-controlled g-factor modulation has now been confirmed 
experimentally~\cite{Salis}.

      \section{Two-Qubit Gates}           %
\label{sec2bit}
The main component for every computer concept
 is a multi-(qu)bit gate,
 which eventually allows calculations
 through combination of several (qu)bits.
Since two-qubit gates are
 (in combination with single-qubit operations)
 sufficient for quantum computation~\cite{DiVincenzo95}---they
form a universal set---we now focus on a mechanism that couples pairs of
spin-qubits. Such a mechanism exists in coupled quantum dots,
 resulting from the combined action of the Coulomb interaction and
 the Pauli exclusion principle.
Two coupled electrons in absence of a magnetic field
 have a spin-singlet ground state,
 while the first excited state in the presence of strong Coulomb repulsion
 is a spin triplet.
Higher excited states are separated from these two lowest states
 by an energy gap,
 given either by the Coulomb repulsion or the single-particle confinement.
The low-energy dynamics of such a system can be described by the
 effective Heisenberg spin Hamiltonian
\begin{equation}\label{Heisenberg}
H_{\rm s}(t)=J(t)\,\,{\bf S}_1\cdot{\bf S}_2,
\end{equation}
where $J(t)$ denotes the exchange coupling between
the two spins ${\bf S}_{1}$ and ${\bf S}_{2}$,
 i.e.\ the energy difference between the triplet and the singlet.
After a pulse of $J(t)$ with
$\int_0^{\tau_s} dtJ(t)/\hbar = J_0\tau_s/\hbar = \pi$ (mod $2\pi$),
the time evolution
$U(t) = T\exp(i\int_0^t H_{\rm s}(\tau)d\tau/\hbar)$
corresponds to the ``swap'' operator $U_{\rm sw}$,
 whose application leads to an interchange of the
 states in qubit~1 and~2~\cite{Loss97}.
While $U_{\rm sw}$ is not sufficient for quantum computation,
 any of its square roots $\sqrtSwap$, say 
$\sqrtSwap \ket{\phi\chi} = (\ket{\phi\chi}+i\ket{\chi\phi})/(1+i)$,
 turns out to be a {\em universal} quantum
 gate.
Thus, it can be used, together with single-qubit rotations,
 to assemble any quantum algorithm.
This is shown by constructing the known universal gate \XOR~\cite{Barenco},
 through combination of
 $\sqrtSwap$  and
 single-qubit operations $\exp(i\pi S_i^z/2)$,
 applied in the sequence~\cite{Loss97},
\begin{equation}
U_{\rm XOR} = e^{i(\pi/2)S_1^z}e^{-i(\pi/2)S_2^z}\,\sqrtSwap\,
e^{i\pi S_1^z}
\,\sqrtSwap .
\end{equation}

With these universal gates at hand, we can reduce the
 study of general quantum computation
 to the study of  single-spin rotations (see Sec.~\ref{secSSpinRot})
 and the {\it exchange mechanism}, in particular
 how $J(t)$ can be controlled experimentally.
The central idea is that $J(t)$ can be switched by raising or lowering
 the tunneling barrier between the dots.
In the following, we shall review our detailed calculations to describe
such a mechanism.
We note that the same principles can also be applied
 to other spin systems in quantum-confined structures, such as
 coupled atoms in a crystal, supramolecular structures,
 and overlapping shallow donors in semiconductors~\cite{Kane,Yablonovitch}
etc., using similar methods as explained below.

\subsection{Coupled Quantum Dots}
\label{ssecldots}
We consider a system of two tunnel-coupled quantum dots, achieved by
gating a two-dimensional electron gas (2DEG), 
as described in Sec.~\ref{ssecQC}.
The dots are arranged in a plane (see Fig.~\ref{doub_dot}), 
at a sufficiently small distance $2a$,
such that the electrons can tunnel between the dots. 
The tunnel junction between the dots, as well as the number of electrons on 
each dot, is controlled by the depleting gates (see Fig.~\ref{figArray}).
We consider the case of similar dots, and each dot contains an 
odd number of electrons with a spin 1/2 ground state. 
Furthermore, we simplify our consideration by 
retaining only one electron per dot and assuming that the 
rest of the electrons form a closed shell and merely contribute to the 
confining potential of the gates. 
We model the two coupled dots with the Hamiltonian~\cite{BLD}
$H = \sum_{i=1,2} h_i+C+H_{\rm Z} = H_{\rm orb} + H_{\rm Z}$,
where the single-electron dynamics in the 2DEG ($xy$-plane) is described
through
\begin{equation} \label{eqnHOneE}
h_i = \frac{1}{2m}\left({\bf p}_i-\frac{e}{c}{\bf A}({\bf r}_i)
\right)^2+W({\bf r}_i),
\end{equation}
with $m$ being the effective mass and $W({\bf r}_i)$ the confinement
potential
as given below.
A magnetic field ${\bf B}= (0,0,B)$ is applied along
 the $z$-axis,
 which couples to the electron spin through the Zeeman interaction $H_{\rm
Z}$
 and to the charge through the vector potential
 ${\bf A}({\bf r}) = \frac{B}{2}(-y,x,0)$.
In almost depleted regions, like few-electron quantum dots,
 the screening length $\lambda$ can be expected to be much larger
 than the screening length in bulk 2DEG regions
 (where it is $40\:{\rm nm}$ for GaAs).
Thus, for small quantum dots, say $\lambda \gg 2a \approx 40\:{\rm nm}$,
we need to consider the bare Coulomb interaction
$C={{e^2}/{\kappa | {\bf r}_1-{\bf r}_2|}}$,
 where $\kappa$ is the static dielectric constant.
The confinement and tunnel-coupling in Eq.~(\ref{eqnHOneE})
 for laterally aligned dots is modeled by the quartic potential
\begin{equation}
\label{eqnV}
W(x,y)=\frac{m\omega_0^2}{2}\left[\frac{1}{4 a^2}\left(x^2-a^2
\right)^2+y^2\right],
\end{equation}
 with the inter-dot distance $2a$ and
 $a_{\rm B}=\sqrt{\hbar/m\omega_0}$
 the effective Bohr radius of the dot.
Separated dots ($a\gg a_{\rm B}$) are thus modeled
 as two harmonic wells with frequency $\omega_0$.
This is motivated by the experimental evidence that
 the low-energy spectrum of single dots is well described
 by a parabolic confinement potential~\cite{tarucha}.

\begin{figure}
\centerline{\psfig{file=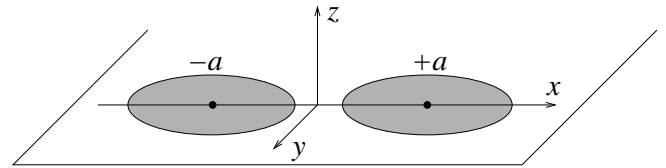,width=8.6cm}}
\caption{\label{doub_dot} Double dot geometry}
\end{figure}

Now we first classify the two-particle states according to the available
symmetries, and then we calculate the phenomenological parameters within 
our toy model.  
The wave function of the two electrons in the double dot (DD) can be chosen
to be a product of an orbital and a spin part. 
Then, in a singlet (triplet) 
state the orbital part is symmetric (antisymmetric) with respect to the 
interchange of the electrons, while the spin part is antisymmetric 
(symmetric).  
With respect to the {\em mirror reflection} in the $yz$-plane~\cite{note1} 
(see Fig.~\ref{doub_dot})
the single-particle states fall into two symmetries, which we label by the
quantum number $n=\pm$. 
The energy difference between the state with $n=-$ and that with $n=+$
is given by $2t_0>0$ ($t_0$ being the interdot hopping amplitude).
Within the low energy sector, the lowest 
singlet state and the triplet states are then given by~\cite{note2}
\begin{eqnarray}\label{states1}
&&|00\rangle=\frac{1}{\sqrt{1+\phi^2}}(d_{+\uparrow}^{\dag}
d_{+\downarrow}^{\dag}-\phi d_{-\uparrow}^{\dag}
d_{-\downarrow}^{\dag})|0\rangle\;,
\nonumber\\
&&|11\rangle=d_{-\uparrow}^{\dag}d_{+\uparrow}^{\dag}|0\rangle\;,\;\;\;\;
|1-1\rangle=d_{-\downarrow}^{\dag}d_{+\downarrow}^{\dag}|0\rangle\;,\\
&&|10\rangle=\frac{1}{\sqrt{2}}(d_{-\uparrow}^{\dag}
d_{+\downarrow}^{\dag}+d_{-\downarrow}^{\dag}
d_{+\uparrow}^{\dag})|0\rangle\nonumber\;,
\end{eqnarray}
where the notation $|SS_z\rangle$ stands for the angular momentum 
representation of the total spin of the two electrons. 
Here, the second quantized operator
$d_{n\sigma}^{\dag}$ creates an electron in the orbital state $n$ with
spin $\sigma$. The vacuum state $|0\rangle$ includes the disregarded
electrons. The {\em interaction} parameter $\phi$ depends on the interplay 
between
the tunneling and Coulomb interaction. We have calculated $\phi$~\cite{VGDL}
within the Hund-Mulliken method,
\begin{equation}
\phi=\sqrt{1+\left(\frac{4t_H}{U_H}\right)^2}-\frac{4t_H}{U_H}\;,
\label{phi1}
\end{equation}
where $t_H$ and $U_H$ are the extended inter-dot tunneling amplitude
and on-site Coulomb repulsion, respectively~\cite{BLD}. We note that
$t_H=t_0+t_C$, with the contribution $t_C$ coming from the Coulomb 
interaction~\cite{BLD} and vanishing with vanishing $t_0$. 
For detached dots we have $\phi=1$, and $\phi<1$
occurs due to double occupancies in
the dots, and $\phi\to 0$ for vanishing Coulomb interaction.
By varying $a$, we plot $\phi$ versus $t_0$ on Fig.~\ref{figphi}.
Next, we note that the singlet state in (\ref{states1}) represents an entangled
state of two electrons, with the entanglement being
\begin{equation}\label{eta}  
\eta=\frac{2\phi}{1+\phi^2}.
\end{equation} 
Here, we used the measure of entanglement introduced 
in Ref.~\cite{Schliemann}.

\begin{figure}
\centerline{\psfig{file=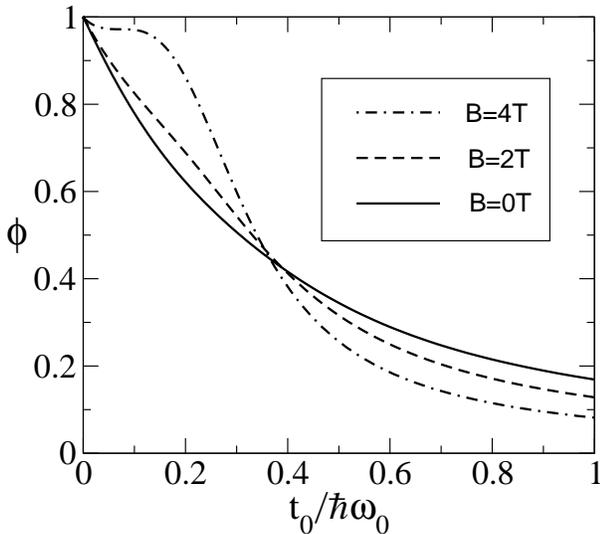,width=8.cm}}
\caption{\label{figphi} Parameter $\phi$ versus the interdot hopping 
amplitude $t_0$ at different magnetic fields (Hund-Mulliken calculation). 
We used GaAs quantum dots with confinement energy $\hbar\omega_0=3\,{\rm meV}$ 
and dielectric constant $\kappa=13.1$.}
\end{figure}
\noindent

Next, assuming adiabatic switching of the coupling constants, 
we arrive at the effective Hamiltonian (\ref{Heisenberg}) by means of
the following mapping~\cite{PG,VGDL}
\begin{eqnarray}\label{mapping}
&&\sum_{\sigma\sigma'}d_{n\sigma}^{\dag}
{\bbox{\sigma}}_{\sigma\sigma'}d_{n'\sigma'}={\bf S}_+\delta_{nn'}+
\left(\frac{\phi_+}{2}{\bf
S}_-+in\phi_-{\bf T}\right)\delta_{-nn'},\nonumber
\end{eqnarray}
\vspace{-15pt}
\begin{eqnarray}
&&\sum_{\sigma}d_{n\sigma}^{\dag}d_{n'\sigma}
=\delta_{nn'}\left[1-\frac{n}{2}\phi_+\phi_{-}\left(
{\bf S}_1\cdotp{\bf S}_2-\frac{1}{4}\right)\right],\;\;\;\;
\end{eqnarray}
where ${\bf S}_{\pm}={\bf S}_1\pm{\bf S}_2$, 
${\bf T}={\bf S}_1\times{\bf S}_2$, 
$\phi_\pm=\sqrt{2}(1\pm\phi)/\sqrt{(1+\phi^2)}$, and $\bbox{\sigma}$ are 
the Pauli matrices. 
Mapping (\ref{mapping}) projects out the higher energy sector of the DD 
and keeps only the states (\ref{states1}).
We note that the spin 1/2 operators ${\bf S}_{1,2}$ are nothing but the
intermixed electron spins, and hence represent the spin degrees of freedom 
of the DD. When detaching the two dots adiabatically, one always obtains
one electron in each dot~\cite{Schliemann} 
and the spins ${\bf S}_{1,2}$ then stand for the true
electron spins. Also note that during an adiabatic coupling of the two
qubits (dots) each of the spins ${\bf S}_{1,2}$ 
carries the initial information of its qubit, which gets modified only through
the Heisenberg exchange interaction $J$, see Eq.~(\ref{Heisenberg}).
The adiabaticity criterion discussed in Sec.~\ref{ssecPulse}
applies hence here with $\delta\varepsilon=\min(\hbar\omega_0,U_H)$. 

The Heisenberg exchange interaction $J$, which is defined as the energy 
difference between the triplet and singlet states (\ref{states1}), 
is the only parameter of interest 
for the two-qubit dynamics (provided the adiabaticity criterion is fulfilled). 
Although for a real structure it is best to have methods to measure $J$ for
different values of the gate voltages, {\em i.e.} to characterize the 
structure experimentally (see Sec.~\ref{secKondo}), we still find it 
instructive to analyse various contributions to $J$ within our realistic model 
of the DD. In particular, we show that breaking the time-reversal symmetry
by means of applying a magnetic field leads to a singlet-triplet transition 
in the DD.
We calculate $J$ using different methods and compare the results.
A generic expression for $J$ is straightforwardly obtained from 
the states (\ref{states1}),
\begin{equation}
J=V+\frac{1-\phi^2}{1+\phi^2}2t_H-\frac{(1-\phi)^2}{1+\phi^2}\frac{U_H}{2}\;,
\label{Jgeneral}
\end{equation}
with $V=\langle T|C|T\rangle-\langle 00^+|C|00^+\rangle$, 
$t_H=t-\langle 00^+|C|00^-\rangle/2$, 
and $U_H=\langle 00^-|C|00^-\rangle-\langle 00^+|C|00^+\rangle$.
Here, $|T\rangle$ stands for any of the triplet states in (\ref{states1}), and
$|00^\pm\rangle$ denotes the singlet state of (\ref{states1}) taken at 
$\phi=\pm 1$. In the Hund-Mulliken approach, $\phi$ is given by (\ref{phi1})
and expressions for  $V$, $t_H$, and $U_H$ were obtained in 
Ref.~\onlinecite{BLD}. The Heisenberg exchange interaction then reads
\begin{equation}
J=V-\frac{U_H}{2}+\frac{1}{2}\sqrt{U_H^2+16t_H^2}\;,
\label{JHundMulliken}
\end{equation}
We note that the component $V$ is responsible for making $J$ ferromagnetic, 
{\em i.e.} $J<0$. 
In the standard Hubbard approach for short-range Coulomb interaction 
it is assumed that there is no overlap between the electron wave function 
on different dots, though there is a finite hopping amplitude $t$ 
(chain model). The exchange terms of the Coulomb interaction hence vanish,
and this corresponds to setting
$V\to 0$, $t_H\to t$, and $U_H\to U$, with $U$ being the 
on-site (short-range) Coulomb repulsion~\cite{BLD}.
The parameter $\phi$ is given by a formula analogous to (\ref{phi1}).
The Heisenberg exchange interaction is then always antiferromagnetic,
$J=\sqrt{(U/2)^2+4t^2}-U/2>0$. 
Finally, in the Heitler-London approach,
the double occupancy of each dot is neglected, which corresponds to setting
$\phi=(1-S)/(1+S)$, where $S$ is the overlap integral 
between the electron wave functions on the two dots. 
Note that the Heitler-London method 
gives qualitatively wrong results 
for both the case of strong and weak Coulomb interaction, since the
interaction parameter $\phi$ is not sensitive to the Coulomb interaction. 
However, for intermediate strengths of the Coulomb interaction 
($U_H\sim \hbar\omega_0$) 
it gives 
qualitatively correct results. The parameters $V$, $t_H$, and $U_H$ 
are identical for both Heitler-London and Hund-Mulliken methods.
We find it appropriate to use the Heitler-London approximation 
for simplicity in a GaAs system. 
Formula (\ref{Jgeneral}) then reduces to~\cite{BLD},
\begin{eqnarray}\label{J}
J &=& \frac{\hbar\omega_0}{\sinh\left(2d^2\,\frac{2b-1}{b}\right)}
\Bigg\{
\frac{3}{4b}\left(1+bd^2\right)\\ \nonumber
 && + c\sqrt{b} \left[e^{-bd^2} \, I_0\left(bd^2\right)
- e^{d^2 (b-1)/b}\, I_0\left(d^2\,\frac{b-1}{b}\right)\right]
\Bigg\},
\end{eqnarray}
where we have introduced the dimensionless distance $d=a/a_{\rm B}$
 between the dots
 and the magnetic compression factor
$b=B/B_0=\sqrt{1+\omega_L^2/\omega_0^2}$
 with the Larmor frequency $\omega_L=eB/2mc$.
The zeroth order Bessel function is denoted by $I_0$.
In Eq.~(\ref{J}),
 the first term comes from the confinement potential,
 while the terms proportional to the parameter
 $c=\sqrt{\pi/2}(e^2/\kappa a_{\rm B})/\hbar\omega_0$ result from
 the Coulomb interaction $C$;
 the exchange term is recognized by its negative sign.
We plot $J$ [Eq.~(\ref{J})] in Fig.~\ref{Jplots} as a function of $B$ and $d$.
We note that $J(B\!=\!0)>0$, which is generally true for a
 two-particle system with time-reversal invariance.
We observe that over a wide range of the parameters $c$ and $a$,
 the sign of $J(B)$ changes from positive to negative
 at a finite value of $B$
 (for the parameters chosen in Fig.~\ref{Jplots}(a) at $B\approx1.3\:{\rm
T}$).
$J$~is suppressed exponentially either
 by compression of the electron orbitals through large magnetic fields
 ($b\gg 1$),
 or by large distances between the dots ($d\gg1$),
 where in both cases the orbital overlap of the two dots is reduced.
This exponential suppression, contained in the $1/\sinh$ prefactor
 in Eq.~(\ref{J}),
 is partly compensated by the exponentially growing
 exchange term $\propto\exp(2d^2(b-1/b))$.
In total, $J$ decays exponentially as $\exp(-2d^2b)$ for large $b$ or $d$.
Since the sign reversal of $J$---signalling a
singlet-triplet crossing---results
from the long-range Coulomb interaction,
 it is not contained in the standard Hubbard model
 which takes only short-range interaction into account.
Fig.~\ref{Jplots} compares the results of different methods.

We remark again that the exponential suppression
 of  $J$ is very desirable for minimizing gate errors. 
In the absence of tunneling between
the dots we still might have direct Coulomb interaction left between
the electrons. However, this has no effect on the spins (qubit)
provided the spin-orbit coupling is sufficiently small, which
is the case for s-wave electrons in GaAs structures with unbroken
inversion symmetry (this would not be so for hole-doped systems
since the hole has a much stronger spin-orbit coupling due to its
p-wave character). For a detailed discussion of spin-orbit interaction
and its effect on exchange and XOR gates we refer to 
Ref.~\onlinecite{cancellation}.

\begin{figure}
\centerline{\psfig{file=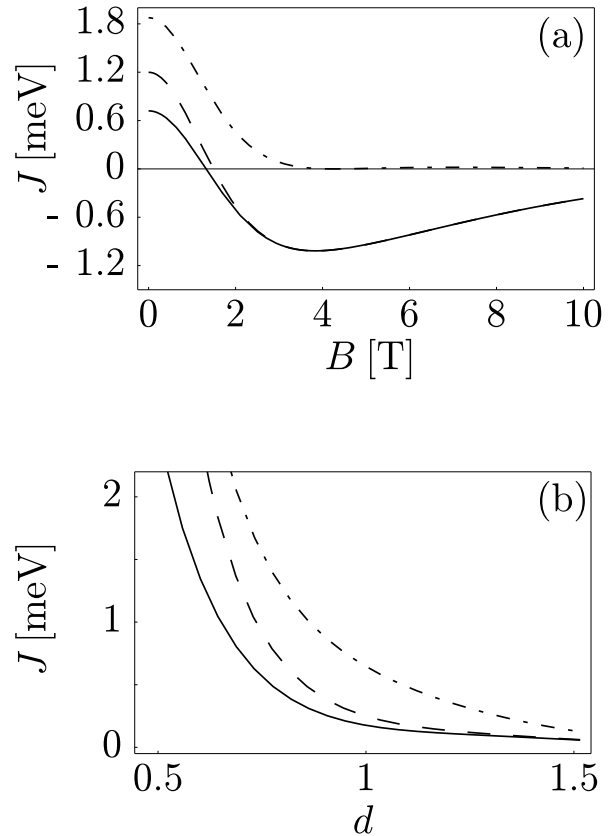,width=8.cm}}
\caption{ \label{Jplots} Exchange
coupling $J$ (full line) for GaAs quantum dots with
confinement energy $\hbar\omega=3\,{\rm meV}$ and $c=2.42$.
For comparison we plot the usual short-range Hubbard result
$J=4t^2/U$ (dashed-dotted line) and the extended Hubbard
result~\protect\cite{BLD}
$J=V+4t_H^2/U_H$ (dashed line).
In (a), $J$ is plotted as a
function of the magnetic field $B$ at fixed inter-dot distance
$d=a/a_{\rm B}=0.7$,
while in (b) as a function of the
inter-dot distance $d=a/a_{\rm B}$ at $B=0$. }
\end{figure}

             \section{Measuring a Single Spin (Read-Out)}        %
\label{secSpinMeas}

\subsection{Spin Measurements through Spontaneous Magnetization}
\label{ssecPM}
One scheme for reading out the spin of an electron on a quantum dot
 is implemented by tunneling of this electron
 into a supercooled paramagnetic dot~\cite{Loss97}.
There the spin induces  a  magnetization nucleation
 from the paramagnetic metastable phase into a ferromagnetic domain,
 whose magnetization direction $(\theta,\varphi)$ is along the measured
spin direction and which
 can be measured by conventional means.
Since this direction is continuous rather than only one of two values,
 we describe this generalized measurement
 in the formalism of positive-operator-valued (POV)
measurements~\cite{Peres}
 as projection into the overcomplete set of spin-$1/2$ coherent states
 $\ket{\theta,\varphi}=\cos(\theta/2)\spup + e^{i\varphi}\sin(\theta/2)
\spdown$.
Thus if we interpret a magnetization direction in the upper hemisphere
 as $\spup$,
 we have a $75$\%-reliable measurement,
 since
 $(1/2\pi)\int_{\theta\geq \pi/2}d\Omega\,
  |\bra{\uparrow\!}\theta,\varphi\rangle|^2 = 3/4$,
 using the normalization constant $2\pi$ for the coherent spin states.

\subsection{Quantum Dot as Spin Filter and Read-Out/Memory Device}
\label{ssecSpinFilter}

We discuss now a setup---quantum dot attached to in- and outgoing
current leads $l=1,\,2$---which
 can be operated as a spin filter,
 or as a read-out  device, or as a spin-memory where a single spin
stores the
information~\cite{Recher}.

A new feature of this proposal is that the spin-degeneracy
 is lifted with {\it different} Zeeman splittings in the dot and
 in the leads,
 e.g.\ by using materials with different effective
g-factors for leads
and dot~\cite{Recher}.
This results in Coulomb
 blockade peaks and spin-polarized currents which are uniquely associated
with the spin state
 on the dot.

The setup is described by a standard tunneling  Hamiltonian
$H_0+H_T$~\cite{Mahan},
 where $H_0=H_L+H_D$ describes the leads and the dot.
$H_D$ includes
 the charging and interaction energies of the electrons in the dot
 as well as their Zeeman energy $\pm g\mu_B B/2$
 in an external magnetic field ${\bf B}$.
The tunneling between leads and the dot is described by
$H_T=\sum_{l,k,p,s}t_{lp}c_{lks}^{\dag}d_{ps}+{\rm h.c.}$,
 where $c_{lks}$ annihilates electrons with spin $s$ and momentum $k$ in 
lead~$l$,
and $d_{ps}$ annihilates electrons in the dot. We consider
the Coulomb blockade regime~\cite{kouwenhoven} where the charge on the dot
is quantized.
Then we
 apply a standard master-equation approach~\cite{cb,Recher}
 with a reduced density matrix of the dot
 and calculate the transition rates in
 a ``golden-rule'' approach up to 2nd order in $H_T$.
The first-order contribution to the current is
 the sequential tunneling current $I_s$~\cite{kouwenhoven},
 where the number of electrons on the dot fluctuates
 and thus the processes of an electron tunneling from the lead onto the
dot
 and vice versa are allowed by energy conservation.
The second-order contribution is
 the cotunneling current $I_c$~\cite{averinnazarov},
 involving
 a virtual intermediate state with a different number of electrons on
 the dot.

We now consider a system,
 where the Zeeman splitting in the leads is negligible (i.e.\ 
much smaller than the Fermi energy),
while on the dot it is given as $\Delta_z = \mu_B |gB|$.
We assume  a small bias $\Delta\mu = \mu_1-\mu_2 >0$
 between the leads at chemical potential $\mu_{1,\,2}$
 and low temperatures so that $\Delta\mu,\, kT < \delta$,
 where $\delta$ is the characteristic energy-level distance on the dot.
First we consider a quantum dot in the ground state,
 filled with an odd number of electrons with total spin $1/2$,
 which we assume to be $\spup$ and to have energy
$E_\uparrow=0$. If an electron tunnels from the lead onto the dot, a spin
singlet is formed with energy $E_S$,
 while the spin triplets are (usually) excited states with energies
 $E_{T_\pm}$ and $E_{T_0}$.
At the sequential tunneling resonance, $\mu_1>E_S>\mu_2$,
 where the number of electrons on the dot fluctuates between $N$ and
$N+1$,
 and in the regime $E_{T_+}-E_S,\,\Delta_z > \Delta\mu,\, kT$,
 energy conservation only allows ground state transitions.
Thus, spin-up electrons are not allowed to tunnel from lead~$1$
 via the dot into lead~$2$,
 since this would involve virtual states $\ket{T_+}$ and $\spdown$,
 and so we have $I_s(\uparrow)=0$ for sequential tunneling.
However, spin down electrons may pass through the dot in
 the process
 \edot{$\downarrow$}\qdot{$\uparrow$}$_i$~$\to$
 \qdot{$\uparrow\downarrow$}$_f$,
followed by
 \qdot{$\uparrow\downarrow$}$_i$~$\to$
 \qdot{$\uparrow$}\edot{$\downarrow$}$\!_f$.
Here the state of the quantum dot is drawn inside the circle,
 while the states in the leads are drawn to the left and right, {\it resp.},
 of the circle.
This leads to a {\it spin-polarized}
 sequential tunneling current $I_s = I_s(\downarrow)$,
 which we have calculated as~\cite{Recher}
\begin{eqnarray}
&& I_s(\downarrow)/I_0=\theta(\mu_1-E_S)-\theta(\mu_2-E_S), \quad
k_B T<\Delta\mu ,
\label{eqnSmallT} \\
&& I_s(\downarrow)/I_0=
\frac{\Delta\mu}{4k_BT}\cosh^{-2}\left[\frac{E_S-\mu}{2k_BT}\right],
\quad k_BT >\Delta\mu,
\label{eqnLargeT}
\end{eqnarray}
where $\mu = (\mu_1+\mu_2)/2$
 and $I_0=e\gamma_1\gamma_2/(\gamma_1+\gamma_2)$.
Here $\gamma_l=2\pi\nu|A_{lnn'}|^2$ is the tunneling rate
 between lead~$l$ and the dot, and we have introduced the matrix elements
$A_{ln'n}=\sum_{ps}t_{lp}\langle  n'|d_{ps}| n\rangle$.
Similarly, for $N$ even we find $I_s(\downarrow)=0$,
while for $I_s(\uparrow)$ a similar result holds~\cite{Recher} as
 in Eqs.~(\ref{eqnSmallT}), (\ref{eqnLargeT}).

Even though $I_s$ is completely spin-polarized,
 a leakage of current with opposite polarization
 arises through cotunneling processes~\cite{Recher};
still the leakage is small, and the efficiency
for $\Delta_z<|E_{T_+}-E_S|$ for spin filtering in
the sequential regime becomes~\cite{Recher} (for $\gamma_1\sim\gamma_2$)
\begin{equation}
\label{efficiencyST}
I_s(\downarrow)/I_c(\uparrow)\sim
\frac{\Delta_z^2}{(\gamma_1+\gamma_2)\max\{k_BT,\Delta\mu\}},
\end{equation}
 and equivalently for
 $I_s(\uparrow)/I_c(\downarrow)$ at the even-to-odd transition.
In the sequential regime
 we have $\gamma_i< k_{B}T,\Delta\mu$,
 thus, for $k_{B}T,\Delta\mu<\Delta_z$,
 we see that the spin-filtering is very efficient.

We discuss now the opposite case where the leads are fully spin polarized
 with a much smaller Zeeman splitting on the dot~\cite{Recher}.
Such a situation can be realized with magnetic
semiconductors (with effective g-factors reaching
100~\cite{Fiederling}) where spin-injection into GaAs has recently been
demonstrated for the first time\cite{Fiederling,Ohno}.
Another possibility would be to work in the quantum Hall regime
 where spin-polarized edge states are coupled to a quantum
 dot\cite{Sachrajda}.
In this setup the device can be used as read-out for the spin state
on the dot.
Assume now that the spin polarization in both leads is up,
and the ground state of the dot contains an odd
 number of electrons with total spin $1/2$.
Now the leads can provide and absorb only
spin-up electrons. Thus,  a sequential tunneling
current will only be possible if the dot state is $\spdown$ (to form a
singlet with the incoming electron, whereas the triplet is excluded by
energy conservation). Hence,
the current is much larger for the spin on the dot being in $\spdown$
 than it is for $\spup$. Again, there is a small cotunneling leakage
current for the dot-state $\spup$, with a ratio of the two
currents given by Eq.~(\ref{efficiencyST}) with $\Delta_z$ replaced by
$E_{T+}-E_{S}$.
Thus, we can probe (read out) the
  spin-state on the quantum dot by measuring the current
which passes through the dot. Given that the
sequential tunneling current is typically on the order of  $0.1-1$
nA~\cite{kouwenhoven}, we can
estimate the read-out frequency $I/2\pi e$ to be on the order of $0.1-1$ GHz.
Combining this with the initialization and read-in techniques,
i.e.\ ESR pulses
to switch the spin state,
 we have a {\it spin memory} at the ultimate single-spin limit,
 whose relaxation time is just the spin relaxation time $T_1$. This
relaxation time can be expected to be on the order of $100$'s of
nanoseconds~\cite{Kikkawa}, and can be directly measured via the
currents when they switch from high to low due to a spin
flip on the dot~\cite{Recher}. Furthermore, the spin decoherence
time $T_2$ can also be measured via the current, if an ESR field is 
applied to the dot in either sequential tunneling or cotunneling 
regime with normal (unpolarized) leads, as shown in Ref.~\onlinecite{EL}.

\section{Accessing the singlet-triplet splitting in double dots} 
                                                                 %
\label{secKondo}
Transport measurements can be used to characterize a quantum dot system 
experimentally.
A main parameter of interest for quantum computing is the exchange 
interaction $J$ between the spins of two neighboring dots. 
We have considered a setup\cite{VGDL}, consisting of two lateral quantum 
dots connected in series between two metallic leads, see Fig.~\ref{figDD}.  
A magnetic field $B$, applied perpendicular to the 
plane of the dots, is used to tune the exchange interaction $J$.
A common gate (not shown), with the gate voltage $V_g$, can be used 
to change the electron occupation number of the double dot (DD). 
The conductance of the DD versus $V_g$ shows peaks of sequential tunneling 
separated by Coulomb blockade valleys [here we consider temperatures smaller 
than the Coulomb correlation energy]. 
We focus on the valley with two electrons in the DD.
Our consideration also holds for a larger ocupation number $M=2N$, with $N$
being odd, provided $N-1$ electrons on each dot form a closed shell and can 
be disregarded. The two (outer shell) electrons are confined by the DD 
potential, however their spin degrees of freedom can be correlated on 
a much smaller energy scale $J$.
Our aim is to provide ways of accessing the exchange interaction $J$ between 
the spins of the two electrons in the DD.

Readily the differential cotunneling
conductance through the DD shows distinct features 
(steps at the bias $\Delta\mu=\pm J$), which allow one to
measure $J$ experimentally.
However, attaching leads to the DD shifts the energy levels, and hence,
modifies $J$. Moreover, measuring a small value of $J$ requires low
temperatures at which the Kondo correlations in the leads may be important.
We show that such Kondo correlations also introduce a correction to $J$, which
is temperature dependent. 
We find that the peculiar features in transport properties are better
pronounced in the Kondo regime. For example, the linear conductance of the DD 
as function of temperature shows a maximum at temperature $T\simeq J>0$, 
which is pronounced only in the Kondo regime. This maximum can be used 
as an alternative way of measuring $J$, having the advantage that, in the 
linear regime, the DD is not affected by the applied bias.

\begin{figure}
\narrowtext {\epsfxsize=6.cm
\centerline{\epsfbox{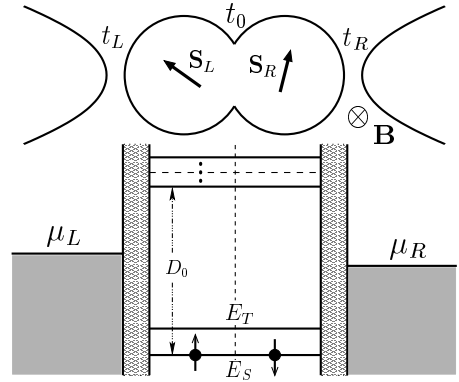}}} \caption{Double-dot 
system containing two electrons and being coupled in series to two
metallic leads at chemical potentials $\mu_R$ and $\mu_L$ with
bias $\Delta\mu=\mu_L-\mu_R$. The electron spins ${\bf S}_L$,
${\bf S}_R$ interact via the exchange interaction $J=E_T-E_S$,
where $E_{T,S}$ is the triplet/singlet energy.} 
\label{figDD}
\end{figure}
\noindent

\subsection{Cotunneling through two tunnel-coupled quantum dots}
\label{ssecCotDD}
In the Coulomb blockade regime, the fluctuations of the number of electrons 
on the DD are strongly suppressed by the Coulomb blockade gap. 
The conductance through the DD is dominated by processes with a virtual 
occupation of the DD by a lead electron (hole), {\em i.e.} cotunneling 
processes. We consider a realistic DD with long range Coulomb interaction
between the two electrons on the DD, as discussed in Sec.~\ref{ssecldots}.
Using a Hund-Mulliken approach, we write down the 
one-electron states as
$\psi_{\pm,\sigma}=\chi_{\sigma}(\varphi_{-a}\pm\varphi_{+a})/\sqrt{2(1\pm S)}$
, where $\chi_{\sigma}$ is the spinor,
$\varphi_{\pm a}$ are the lowest orbitals of single dots situated 
at $x=\pm a$, and
$S=\langle\varphi_{\pm a}|\varphi_{\mp a}\rangle$ is the overlap integral. 
The lowest in energy singlet and triplet are given then by Eq.~(\ref{states1}),
and the interaction parameter $\phi$ is calculated according to 
(\ref{phi1}).
The attached leads are described by  
$H_l=\sum_{\alpha k\sigma}\varepsilon_kc_{\alpha k\sigma}^{\dag}
c_{\alpha k\sigma}$,
where $c_{\alpha k\sigma}^{\dag}$ creates an electron with
momentum $k$ and spin $\sigma$ in lead $\alpha=L,R$.
The tunneling between the DD and the leads is described by
$H_T=\sum_{n\alpha k\sigma}(t_{\alpha n}
c_{\alpha k\sigma}^{\dag}d_{n\sigma}+\mbox{h.c.})$. Here,
$d_{n\sigma}$ annihilates an electron in the state $\psi_{n\sigma}$.
The tunneling amplitudes are given by
$t_{L,\pm}=t_L/\sqrt{2(1\pm S)}$ and $t_{R,\pm}=\pm t_R/\sqrt{2(1\pm S)}$,
with $t_\alpha$ being the apmlitude to tunnel from lead $\alpha$ onto the
the adjusting dot at $t_0=0$. 
We map our problem onto a two-level system, with level 1 corresponding to
the singlet state and level 2 to the three triplet states. 
The occupation probabilities of these two levels, $\rho_1$ and $\rho_2$, 
are given by
\begin{equation}
\rho_1=\frac{1}{1+3\exp\left(
-J/T_{\rm eff}
\right)}=1-\rho_2\;.
\label{rho1}
\end{equation}
The effective temperature $T_{\rm eff}$ depends on the applied bias 
$\Delta\mu$, and thus, describes the heating effects on the DD.
Solving a master equation for $\rho_1$
and $\rho_2$ in the cotunneling regime, we find
\begin{equation}
\frac{1}{T_{\rm eff}}=\frac{1}{T}-\frac{1}{J}
\ln\frac{1+\lambda(J/T,\Delta\mu/T)\phi_-^2}
{1+\lambda(-J/T,\Delta\mu/T)\phi_-^2}\;,
\label{theta1}
\end{equation} 
where
$\lambda(u,v)=\frac{1}{4}
\frac{\sinh(v)}{1-\tanh(u/2)}
\frac{\tanh(v/2)-(v/u)\tanh(u/2)}{\cosh(u)-\cosh(v)}$, and the parameter 
$\phi_-$ was introduced in Sec.~\ref{ssecldots}. We note that the 
heating effect depends on the interaction parameter $\phi$, and for 
$\phi\to 1$ it vanishes. Also, the heating effect is pronounced only for biases
$|\Delta\mu|\geq |J|$, and it vanishes at high temperatures. 
For the vicinity of $\Delta\mu=\pm J$,
we define a characteristic temperature of a {\em strong} heating regime, 
given by 
$T_{\rm h}=|J|/w\left(8/\phi_-^{2}\right)$, 
where the function $w(x)$ is defined 
for $x\geq e$ by 
$w(x)=\ln\left(x\ln\left(x...\right)\right)$.
Bellow this temperature $T_{\rm h}$, the exponential dependence of 
$\exp(-J/T_{\rm eff})$ on $T$ is replaced by a power law dependence;
the temperature $T$ competes with $|\Delta\mu|-|J|$, and as the latter
becomes larger, the occupation probabilities cease to depend on $T$.  
For the strong heating regime $T<T_{\rm h}$, we find
\begin{equation}\label{Teff} 
T_{\rm eff}\simeq\frac{|J|}{\ln\left(1+\frac{8}{\phi_-^2}
\frac{|J|}{\max\left(T,|\Delta\mu|-|J|\right)}\right)},\;\;\;\;
|\Delta\mu|\geq |J|.
\end{equation}

The current through the DD consists of an elastic and inelastic component,
$I=I_{\rm el}+I_{\rm inel}$. In the middle of the Coulomb blockade valley 
the current components are given by
\begin{eqnarray}
I_{\rm el}&=&\frac{e}{h}\frac{2\gamma^2}{(1-S^2)^2}\left[\phi_-^2\phi_+^2
\rho_1+8S^2\rho_2\right]\Delta\mu\;,\\
I_{\rm inel}&=&\frac{e}{h}\frac{2\gamma^2\phi_-^2}{1-S^2}
\left\{\left[\Theta(-J+\Delta\mu)-\Theta(-J-\Delta\mu)\right]3\rho_1
\right.\nonumber\\
&&\left.+\left[\Theta(J+\Delta\mu)-\Theta(J-\Delta\mu)\right]\rho_2
\right\}\;,
\end{eqnarray}
where $\Theta(J)=J/(1-\exp(-J/T))$, and 
$\gamma=\pi\nu t_Lt_R/E_C$,
with $E_C$ being the Coulomb blockade half-gap, and $\nu$ the density of 
states in the leads.
In the absence of heating, when 
$T_{\rm eff}=T$, the elastic component $I_{\rm el}$ is linear in the 
applied bias $\Delta\mu$, 
and the inelastic one $I_{\rm inel}$ exhibits a threshold-like switching-on at 
$|\Delta\mu|=|J|$, for $T<|J|$~\cite{VGDL}. This results in steps in 
differential conductance versus $\Delta\mu$ at $|\Delta\mu|=|J|$, which can 
be used to measure the singlet-triplet splitting experimentally~\cite{SDF}.
For $T\ll|J|$, the step height was found to be 3 times larger on the singlet 
side than on the triplet side~\cite{VGDL}.
In the strong heating regime, the differential conductance 
$G=edI/d\Delta\mu$  provides
information also about the DD parameter $\phi$. 
We plot $G(\Delta\mu)$ in Fig.~\ref{figheating}, for (a) a singlet and 
(b) a triplet ground state. 
The dashed line shows the result of a calculation where 
the heating effects are neglected. We find that, in the strong heating 
regime, 
$dI_{\rm inel}/d\Delta\mu$ has
a negative (positive) slope on the plateau $|\Delta\mu|>|J|$ for the ground 
state being a singlet (triplet). The slope of the elastic component is not 
generic, but depends on the interplay between the parameters $S$ and $\phi$.
However, we still find that the slope of the total conductance is not changed
qualitatively by the elastic component over a large range of parameters 
calculated in the Hund-Mulliken method, Sec.~\ref{ssecldots}. 
\begin{figure}
\narrowtext {\epsfxsize=8.cm
\centerline{\epsfbox{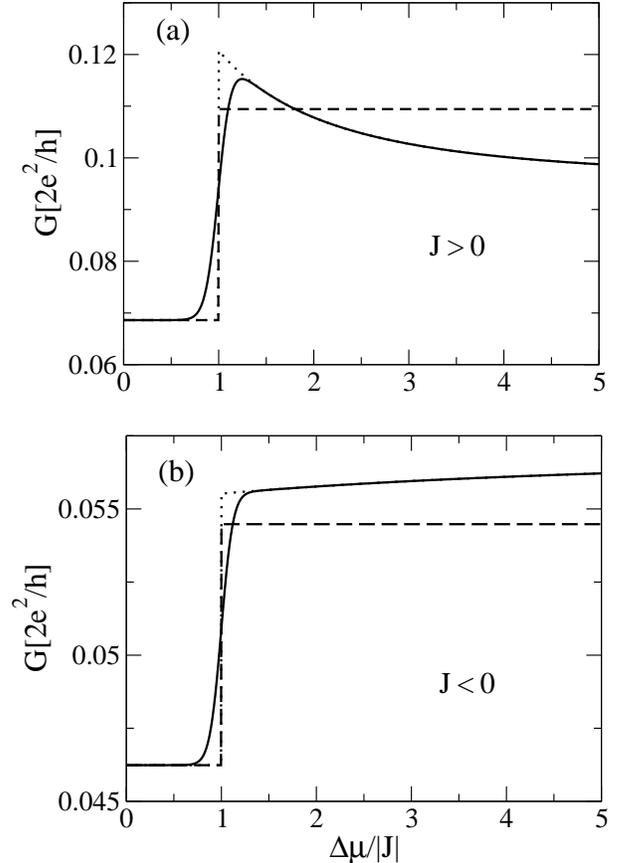}}} 
\caption{Differential conductance $G=dI/d\Delta\mu$ 
versus bias calculated for
dot separation $a/a_B=0.7$ and coupling to the leads given by
$\gamma=0.1$ at a magnetic field: 
(a) $B=0$ and (b) $B=1.5\;{\rm T}$. The solid line and the dotted line
are calculated in the strong heating regime at 
$T=0.2T_{\rm h}$ and $T=0$, respectively. 
The dashed line corresponds to neglecting 
the heating effects, {\em i.e.} $T_{\rm eff}\to T$, and was calculated for 
$T=0$. Thus we see that $G$ is monotonically increasing for $J<0$, while
it has a maximum for $J>0$.} 
\label{figheating}
\end{figure}
\noindent
Next, we consider the case of weakly coupled dots 
($\phi_-^2\ll 1$) and show how
the parameter $\phi$ can be 
extracted from $G(\Delta\mu)$ in the strong heating regime. 
For the singlet ground state, and the bias satisfying
$T\ll |\Delta\mu|-|J|\ll |J|/\phi_-^2$, we find
\begin{equation}
\phi_-^2=\frac{A}{1-A\frac{|\Delta\mu|-|J|}{2|J|}},\;\;\;\;\;\;
A\equiv -\frac{|J|}{\Delta G}\frac{dG}{d\Delta\mu},
\label{phimin}
\end{equation} 
where $\Delta G=G(\Delta\mu)-G(\infty)$. For the triplet ground state
one should replace $\phi_-^2\to{1\over 3}\phi_-^2$ in (\ref{phimin}).
In Fig.~\ref{figheatingA}, we plot the rhs of the equation for $\phi_-^2$
in (\ref{phimin}) at different temperatures. 
In the strong heating regime ($T\ll T_{\rm h}$), the curve saturates
at the value of $\phi_-^2$ with increasing bias. 
This behavior can be used to measure the interaction parameter $\phi$ 
experimentally.
We note that one can reliably determine
$G(\infty)$ experimentally, as the conductance at
$\Delta\mu\gg |J|/\phi_-^2$,
only if the sequential tunneling processes can be excluded. 
For the middle of
the Coulomb blockade valley, we require $|J|+|\Delta\mu|<E_C$, to avoid
sequential tunneling via the {\em heated} excited state~\cite{WegNaz}.
Furthermore, we note that at low temperatures Kondo resonances can develop
at $\Delta\mu=\pm J$, invalidating Eq.~(\ref{phimin}). 
However, one can avoid this problem by satisfying either $T\gg T_K$ or 
$|\Delta\mu|-|J|\gg T_K$, with $T_K$ being the energy scale of
the Kondo resonance.

\begin{figure}
\narrowtext {\epsfxsize=8.cm
\centerline{\epsfbox{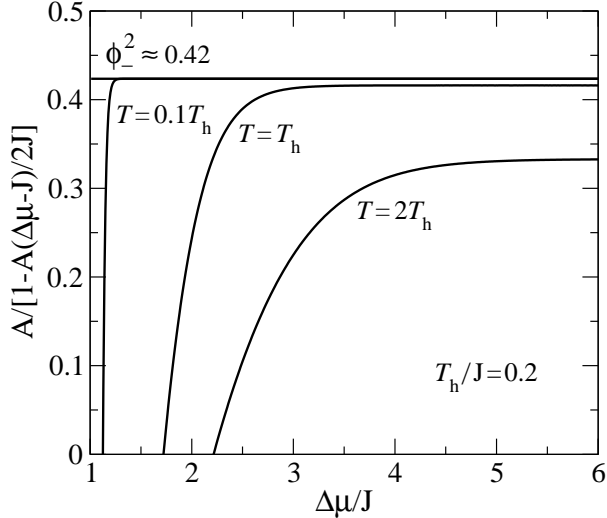}}} 
\caption{A way to measure $\phi_-^2$ experimentally, see Eq.~(\ref{phimin}).
For the calculation we used a dot separation of $a/a_B=1$ and $B=0$ 
(singlet ground state). Note that at $T>T_{\rm h}$ the lines saturate at a 
value which differs strongly from $\phi_-^2$. 
} 
\label{figheatingA}
\end{figure}
\noindent

\subsection{Kondo effect of two coupled dots at the 
singlet-triplet degeneracy point}

Attaching metallic leads to a DD can give rise to Kondo 
correlations at low temperatures. On the one hand, such correlations can 
enhance the conductance through a Coulomb-blockaded DD, 
making the transport measurements more accessible. 
On the other hand, Kondo correlations can modify the studied system 
and introduce
a discrepancy between the measured values and the {\em bare} values of the
system parameters. Knowledge of the Kondo effect in double dots would allow 
one to optimize the experimental setup, in order to obtain reliable data.

We adopted a ``poor man's'' scaling approach~\cite{AndersonHewson} 
to the DD system on
Fig.~\ref{figDD} and obtained an effective Hamiltonian~\cite{VGDL}, 
describing the flow (with lowering $T$) into the Kondo regime of a DD at the 
singlet-triplet degeneracy point $J=0$. We found that the 4-fold 
degeneracy of the DD enhances the Kondo correlations on the Fermi surface, 
as compared to the case when the dots are detached from each 
other, or when the DD spins are locked into a spin 1 (triplet side).
The Kondo temperature at $J=0$ is given by~\cite{VGDL} 
$T_K=D_0\exp\left(-\eta/\nu I_0\right)$, 
where $\eta\leq 0.5$ is a non-universal
number dependent on the internal features of the DD, $D_0\simeq\hbar\omega_0$
is the cutoff energy (see Fig.~\ref{figDD}), and $I_0=(t^2_L+t^2_R)/E_C$.
We find that, at temperatures $T\lesssim T_K$, the DD undergoes a 
strong renormalization of its energy levels, resulting in a flow of the 
exchange interaction $J$~\cite{VGDL}. 
Thus, we conclude that, at such temperatures and at
$J\lesssim T_K$, 
the Kondo correlations strongly modify the coupling 
constant $J$, making any direct experimental measurement of the bare value of 
$J$ hardly possible. However, at larger values of $J>0$, one can make use of 
the Kondo correlations and still have a reliable measurement of $J$. 

We calculated the current through the DD at a bias $\Delta\mu$, 
and found the renormalization of the linear conductance 
$G=\left.\left(dI/d\Delta\mu\right)\right|_{\Delta\mu=0}$ due to Kondo correlations. 
In Fig.~\ref{figKondo} we plot the linear $G$ versus the inter-dot
tunneling amplitude $t_0$ for different values of the magnetic field $B$.
At $B=0$, the renormalized conductance
(solid line) shows a sharp peak at a small 
value of $t_0$, owing to a competition between the Kondo effect of each dot 
with the adjusting lead and the antiferromagnetic exchange $J$. The peak
position corresponds to $J(t)\simeq T_K$~\cite{Izumida}.  At larger values
of $t_0$, a second broader peak occurs, which is sensitive to applying a weak 
magnetic field, such that $J>0$. We find that the broad peak is present only 
if $J$ deviates from $4t_H^2/U_H$ by the contribution $V$ 
(see Sec.~\ref{ssecldots}), which comes from 
the long range Coulomb interaction~\cite{BLD} 
(compare the solid lines with the dot-dashed line in Fig.~\ref{figKondo}).
Note that exactly this contribution to $J$ is responsible for the
singlet-triplet transition in DDs. Thus, we have shown that the long range 
part of the Coulomb interaction can be probed experimentally in DDs, 
and screening effects can be studied.

On the left inset of Fig.~\ref{figKondo}, we plot the temperature 
dependence of the linear conductance calculated with taking into account the 
Kondo correlations (solid line) and neglecting them (dotted line), the 
latter corresponding to the cotunneling calculation of 
Sec.~\ref{ssecCotDD}.
For the case with Kondo correlations, we
find a pronounced maximum in the linear $G$ versus $T$ at $T\simeq J$, 
which can be used to estimate $J$ experimentally. 

The $B$ dependence of the linear $G$ shows a peak
at the singlet-triplet transition, which grows with lowering $T$ down to
$T_K$~\cite{VGDL}, 
see right inset of Fig.~\ref{figKondo}. Note that the enegry scale
for the Kondo effect on the triplet side ($J<0$) is monotonically decreasing 
with increasing $|J|$~\cite{EtoNaz,PG}, 
as follows from a two stage RG procedure valid on that side. 
Furthermore, the strong coupling limit (not shown in Fig.~\ref{figKondo}) 
occurs in two stages with lowering $T$ on the triplet side, 
resulting first in an increase and then,
at a lower energy scale, in a decrease of the conductance~\cite{PG1}.

\vspace{+5pt}
\begin{figure}
{\epsfxsize=8.5cm
\centerline{\epsfbox{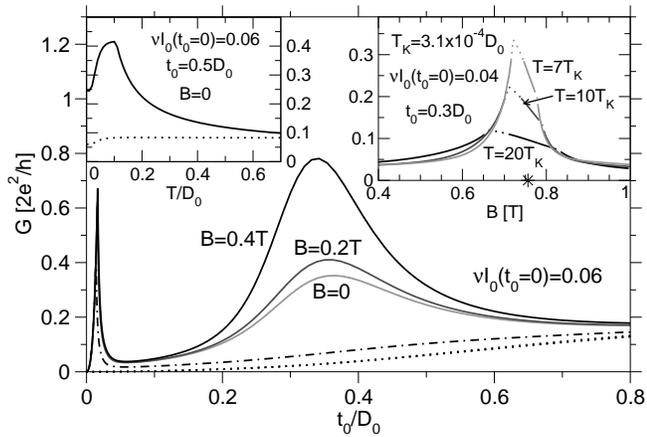}}}\vspace{0pt}
\caption{Linear conductance $G$ at 
different values of $B$. Dotted lines: cotunneling contributions. Dot-dashed 
line: $G$ vs $t_0$ at $B=0.4T$ neglecting the long range part of the Coulomb 
interaction. For definiteness we keep the DD in the middle of 
the Coulomb blockade valley by adjusting the gate voltage $V_g$ 
when varying $t_0$, and choose 
$t_L=t_R$. Left inset: $G$ vs $T$ showing a peak at $T\simeq J$; 
dotted line is the cotunneling contribution.
Right inset: $G$ vs $B$ at the singlet-triplet transition; the kinks in the 
dotted-line regions come from a simplified treatment of the Kondo effect
crossover regions and will be smoothened in an exact treatment; the star 
denotes the value of $B$ at which the singlet-triplet transition occurs at 
high temperatures ($T\gg T_{K}$).}
\label{figKondo}
\end{figure}
\noindent


                      \section{Conclusions}                      %
\label{secConclusions}

We have described a concept for a quantum computer
 based on electron spins in quantum-confined nanostructures, in particular
quantum dots,
 and presented theoretical proposals  for manipulation, coupling
and detection of spins in such structures.
We have discussed the requirements for
coherence, switching times, read-out, gate operations
 and their actual realization.
By putting it all together,
 we have illustrated how a scalable, all-electronically controlled
 quantum computer can be envisioned.

                        \acknowledgements                        %
\addcontentsline{toc}{section}{Acknowledgments}
We acknowledge support from the Swiss NSF, DARPA, and ARO.

\ifpreprintsty\else\end{multicols}\fi


\begin{thebibliography}{}
\bibitem{note1} 
Note that, at $B\neq 0$, this implies also $B\to -B$.
\end{thebibliography}
\end{document}